\documentclass[conference]{IEEEtran}

\usepackage{cite}
\usepackage{amsmath,amssymb,amsfonts}
\usepackage{algorithmic}
\usepackage{graphicx}
\usepackage{textcomp}
\usepackage{xcolor}
\usepackage{booktabs}
\usepackage{url}
\usepackage{listings}
\usepackage{multirow}
\usepackage{array}

\newcommand{\rev}[1]{#1}

\begin{document}

\title{Self-Improving AI Coding Agents Through Accumulated Behavioral Rules: A Closed-Loop Framework}

\author{
\IEEEauthorblockN{Aditya Aggarwal}
\IEEEauthorblockA{\textit{Microsoft} \\
aditya.aggarwal@microsoft.com}
\and
\IEEEauthorblockN{Nahid Farhady Ghalaty}
\IEEEauthorblockA{\textit{Microsoft} \\
nahidf@microsoft.com}
}

\maketitle

\begin{abstract}
LLM-based coding agents repeat the same classes of mistakes across sessions because they lack a mechanism to retain corrections from human review feedback. We present a closed-loop framework in which every accepted review comment is codified as a persistent behavioral rule, progressively expanding the set of error classes the agent can self-detect. The framework combines an accumulating rule set in a version-controlled instruction file, a self-review checklist executed before code submission, and automated validation that ensures rule-set integrity as it grows. In deployment across a 35+ service microservices platform, the rule set grew from 5 to 18 behavioral rules, 15+ language-specific standards, and a 15-item self-review checklist, all derived from real review feedback. We present empirical results from 11 recorded working sessions spanning code generation, PR review, incident investigation, and cross-service refactoring. We observe that accumulated rules shift review effort from low-level correctness toward design-level validation, achieve a measured 0\% recurrence rate for ruled-against error classes, and transfer across heterogeneous agent interfaces. We compare our approach against related work in experiential LLM learning (Reflexion, ExpeL, Voyager) and automated code review (CodeReviewer, SWE-bench agents), showing that our framework achieves persistent cross-session learning without weight updates, operates on production codebases rather than synthetic benchmarks, and addresses an orthogonal dimension (behavioral consistency over time) that existing benchmarks do not measure. The result is a coding agent that improves with every review cycle, accumulating the engineering wisdom of its human collaborators without changing a single model weight.
\end{abstract}


\section{Introduction}

AI coding agents powered by Large Language Models have demonstrated remarkable capability in code generation, bug fixing, and code review~\cite{chen2021evaluating, fan2024llm}. A persistent limitation, however, is their inability to learn from session to session. When a human reviewer catches an error in AI-generated code (say, disposing of a resource whose lifetime is managed by a framework-level factory), the agent will reproduce the same mistake in the next session because it has no mechanism to retain that correction.

Current approaches to improving LLM coding quality fall into three categories: (1)~fine-tuning on curated code datasets, which is expensive and organization-specific; (2)~retrieval-augmented generation (RAG), which surfaces relevant code context but does not encode behavioral constraints; and (3)~prompt engineering with static instructions~\cite{zhou2023leasttomost}, which captures some conventions but does not evolve with experience. None of these implements a feedback loop in which review outcomes automatically strengthen future agent behavior.

We propose a fourth approach, \textit{accumulated behavioral rules}, built on one core insight:

\begin{quote}
\textbf{Every accepted review comment is a self-review rule.} When a reviewer catches something the agent missed, codify the pattern so the agent catches it next time. The goal is zero avoidable review comments.
\end{quote}

This paper describes the architecture of a system that implements this principle, reports on its deployment in a production environment with quantitative empirical results, and positions the contribution against the growing body of work on experiential learning for LLM agents~\cite{li2024review}.

We make the following contributions:

\textbf{C1:} A closed-loop framework that converts accepted human review feedback into persistent behavioral rules, creating a ratchet effect where the set of prevented error classes grows monotonically over time.

\textbf{C2:} An operational memory mechanism for coding agents implemented through version-controlled instruction files rather than weight updates, making it model-agnostic, tool-agnostic, and immediately deployable.

\textbf{C3:} Empirical results from 11 recorded working sessions (code generation, PR review, cross-service refactoring, incident investigation) showing rule accumulation dynamics, error-class suppression rates, and knowledge transfer patterns.

\textbf{C4:} A systematic comparison against related work in experiential LLM learning (Reflexion~\cite{shinn2023reflexion}, ExpeL~\cite{zhao2024expel}, Voyager~\cite{wang2023voyager}) and automated code review (CodeReviewer~\cite{li2022codereviewer}, SWE-agent~\cite{yang2024sweagent}), identifying the orthogonal dimension (persistent behavioral consistency across sessions) that our framework addresses and existing benchmarks do not measure.

\subsection{Scope and Positioning}

To preempt a natural question, ``Is this just disciplined prompt engineering?'', we clarify what the framework is and is not.

\textbf{This framework is:}
\begin{itemize}
\item A persistent organizational memory mechanism for coding-agent behavior, where memory content is derived exclusively from accepted review outcomes.
\item A repeat-error prevention system that converts point corrections into class-level constraints.
\item A shared, version-controlled knowledge artifact that is reusable across models, tools, and team members.
\end{itemize}

\textbf{This framework is not:}
\begin{itemize}
\item Weight updates, model retraining, or reinforcement learning from human feedback (RLHF)~\cite{ouyang2022training}.
\item A replacement for human review. It augments review by eliminating previously-seen error classes, freeing reviewers for higher-order concerns.
\item A proof of semantic correctness; it prevents known failure modes, not all possible failures.
\item A general-purpose memory system. It is scoped to coding-agent behavioral constraints within a defined codebase.
\end{itemize}

The distinction from static prompt engineering is the \textit{feedback loop}. The instruction file is not written once and maintained passively. It grows as a direct function of review outcomes, with each new rule traceable to a specific accepted comment. This makes the system self-improving in a way that static instruction files are not.

\rev{Having framed what the contribution is and is not, we now describe the framework that implements the closed loop, starting with the structured instruction file that holds the accumulated rules.}

\section{Framework Architecture}

\begin{figure}[htbp]
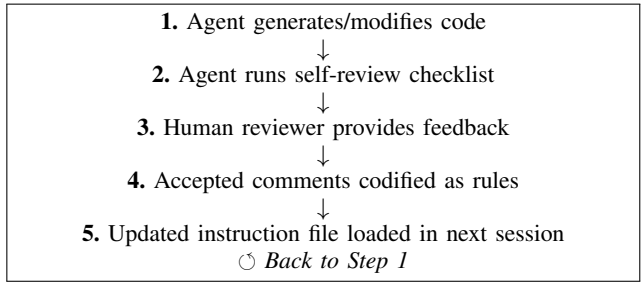

\centering
\fbox{\parbox{0.92\columnwidth}{%
\centering\small
\textbf{1.}~Agent generates/modifies code

$\downarrow$

\textbf{2.}~Agent runs self-review checklist

$\downarrow$

\textbf{3.}~Human reviewer provides feedback

$\downarrow$

\textbf{4.}~Accepted comments codified as rules

$\downarrow$

\textbf{5.}~Updated instruction file loaded in next session

$\circlearrowleft$ \textit{Back to Step 1}
}}
\caption{The closed-loop feedback cycle. Each accepted review comment is codified as a persistent behavioral rule in a version-controlled instruction file, expanding the agent's self-review capability for all subsequent sessions.}
\label{fig:loop}
\end{figure}

The framework comprises several interconnected components. We describe each below, then formalize the rule representation and lifecycle. Fig.~\ref{fig:loop} provides a visual overview of the closed-loop process.

\subsection{The Structured Instruction File}

At the center of the framework is a structured instruction file that LLM-based coding agents load as system context at the start of every session. Several emerging conventions exist for this mechanism (e.g., AGENTS.md, rules files, instruction files), all following the same principle: a version-controlled Markdown file that shapes agent behavior. The file is organized into five sections, though the specific section names and content are domain-dependent. For example, in our deployment:

\begin{enumerate}
\item \textbf{Behavioral Rules:} High-level operational guardrails (e.g., ``Never commit without user approval'').
\item \textbf{Code Standards:} Language-specific rules derived from review feedback (e.g., ``Never use string interpolation in log calls; use message templates'').
\item \textbf{Self-Review Checklist:} A numbered checklist that agents execute before approving any code.
\item \textbf{Anti-Patterns:} Specific patterns the agent is prohibited from generating, with explanations and correct alternatives.
\item \textbf{Workflow Rules:} Automatic triggers that tell agents when to invoke specific tools or sub-agents based on the task type.
\end{enumerate}

The critical property is that the file is append-friendly and monotonically growing. Rules are added and occasionally refined but rarely removed, accumulating the team's collective engineering judgment over time.

\subsection{Rule Representation}

While rules are stored as human-readable Markdown, each rule implicitly conforms to a structured schema:

\begin{lstlisting}
Rule ID:         R17
Category:        Architecture
Trigger Origin:  Human PR reviewer
Scope:           New internal HttpClient registrations
Constraint:      Add client to AllowIPV4Internal
                 SSRF policy
Rationale:       Prevent runtime SSRF policy
                 validation failures
Checklist Map:   Self-review item #9
Added:           2026-02-28
Traced To:       PR review comment on missing
                 SSRF allowlist entry
\end{lstlisting}

This schema serves three purposes: (1)~it makes origin tracking concrete, linking each rule to the review event that produced it; (2)~it maps rules to self-review checklist items, creating a verifiable bridge between the instruction file and the pre-submission check; and (3)~it supports future tooling for conflict detection and rule effectiveness scoring.

\subsection{The Feedback Loop}

The self-improvement cycle operates in five steps:

\begin{enumerate}
\item The agent generates or modifies code in response to a task.
\item The agent executes its self-review checklist before presenting the code.
\item A human reviewer examines the code and provides feedback.
\item For each accepted comment that identifies a generalizable class of mistake (not a one-off typo), a new rule is added to the instruction file.
\item The updated instruction file is loaded in all subsequent sessions, preventing recurrence.
\end{enumerate}

This creates a ratchet effect: the set of error classes the agent can self-detect grows monotonically. The instruction file serves as organizational memory that persists across sessions, models, and tool boundaries.

\subsection{Rule Lifecycle and Governance}

Because the rule set grows monotonically, governance is essential to maintain quality. We address four operational questions:

\textbf{Who decides whether a comment represents a class of mistake versus a one-off issue?} The engineer who receives the review feedback makes this judgment. The heuristic is: ``Would this mistake plausibly recur in a different context?'' If yes, it becomes a rule.

\textbf{When is an existing rule refined instead of adding a new rule?} When a new review comment reveals that an existing rule is too broad or too narrow, the existing rule is updated in place. The rule ID is preserved; the constraint text is sharpened. Section~\ref{sec:conflict} describes a concrete example.

\textbf{How are duplicate or overlapping rules handled?} The instruction file is reviewed periodically for redundancies. When two rules cover overlapping ground, the more specific rule subsumes the general one.

\textbf{Who owns arbitration when two rules conflict?} The instruction file lives in a shared repository where changes require pull requests. Conflicting rules are resolved by the team through the normal review process.

\subsection{Illustrative Rule Examples}

To make the framework concrete, we present five representative rules from the deployed instruction file:

\begin{enumerate}
\item \textbf{Resource lifetime management.} A reviewer caught the agent wrapping a factory-created \texttt{HttpClient} in a \texttt{using} block. The rule states that \texttt{HttpClient} instances from \texttt{IHttpClientFactory} must not be disposed by the caller.
\item \textbf{Structured logging.} A reviewer flagged string interpolation in log calls. The rule requires message templates to preserve structured log properties.
\item \textbf{Boolean logic correctness.} The agent generated an incorrect guard clause using \texttt{!A || !B || !C} (always true by De Morgan's law). The rule requires \texttt{!A \&\& !B \&\& !C} for conjunctive negation.
\item \textbf{Null safety enforcement.} A reviewer caught the null-forgiving \texttt{!} operator. The rule requires \texttt{?? throw new ArgumentException(...)} for explicit validation.
\item \textbf{Security configuration.} The agent registered a new internal HTTP client without adding it to the SSRF allowlist policy. The rule requires every new \texttt{HttpClient} registration include an SSRF policy entry.
\end{enumerate}

\subsection{Workspace Validation as Reliability Mechanism}

As the rule set grows, it becomes a critical organizational artifact. A malformed rule addition could silently degrade the system. The framework includes automated validation that runs at session start:

\begin{itemize}
\item Agent definition files have valid frontmatter and required fields.
\item Skill directories contain properly structured instruction files.
\item Knowledge documents carry freshness dates, flagging stale content.
\item The instruction file contains all required sections.
\item Automation scripts exist and tool-server configurations are well-formed.
\end{itemize}

This validation supports the monotonic-growth principle: the rule set can grow safely because every addition is validated against a known-good schema.

\subsection{Session Continuity}

The framework maintains session continuity through two mechanisms: (1)~a session handoff file capturing in-progress work, pending reviews, and suggested next actions; and (2)~a task log providing a chronological record of completed work. Together, these ensure that accumulated knowledge includes not only behavioral rules but also ongoing work context.

\section{Deployment Experience}

\subsection{Environment}

The framework was deployed in a large-scale microservices platform comprising 35+ services on a managed container orchestration platform with service mesh. The codebase spans approximately 50,000+ lines of shared infrastructure code. LLM-based coding agents were used for code generation, review, architecture investigation, incident response, and cross-service refactoring across multiple agent interfaces, all reading the same instruction file.

\subsection{Rule Accumulation Over Time}

Over a multi-week deployment period, the rule set grew as shown in Table~\ref{tab:composition}.

\begin{table}[htbp]
\caption{Rule Set Composition at End of Observation Period}
\label{tab:composition}
\centering
\begin{tabular}{lr}
\toprule
\textbf{Component} & \textbf{Count} \\
\midrule
Behavioral Rules & 18 \\
Code Standards (typed language) & 15+ \\
Self-Review Checklist Items & 15 \\
Anti-Pattern Rules & 6 \\
Custom Agent Definitions & 13 \\
Operational Skills & 10 \\
Knowledge Documents & 6 \\
\bottomrule
\end{tabular}
\end{table}

\subsection{Knowledge Source Distribution}

Analysis of the first 18 accumulated behavioral rules reveals four distinct sources of learning, as shown in Table~\ref{tab:sources}.

\begin{table}[htbp]
\caption{Knowledge Source Distribution}
\label{tab:sources}
\centering
\begin{tabular}{lrl}
\toprule
\textbf{Source} & \textbf{Count (\%)} & \textbf{Description} \\
\midrule
Human PR reviewers & 7 (39\%) & Domain-specific \\
Automated review bots & 4 (22\%) & Formatting, style \\
Self-discovered & 4 (22\%) & Proactive analysis \\
Production errors & 3 (17\%) & Runtime failures \\
\bottomrule
\end{tabular}
\end{table}

The dominance of human reviewer feedback (39\%) validates the core premise: human review is the highest-quality signal for rule generation.

\subsection{Observed Effects}

\textbf{Suppression of previously-seen error classes:} Across 11 recorded working sessions following rule additions, we observed \rev{no recurrences} of any ruled-against pattern \rev{within the observation window} (see Section~\ref{sec:results}\rev{; this should be read as suppression under the observed deployment conditions rather than a permanent property of the underlying model}).

\textbf{Shift in review focus:} Reviewers reported spending less time on mechanical issues and more on architectural appropriateness and design trade-offs.

\textbf{Cross-interface transfer:} Rules added through one agent interface immediately benefited sessions conducted through a different interface, because all interfaces consumed the same instruction file.

\textbf{Onboarding acceleration:} New team members inherited the full accumulated rule set immediately, bypassing the learning curve that originally produced those rules.

\textbf{Specificity correlated with effectiveness:} Early rules were broad. Later rules were precise. Specific, actionable rules were more reliably followed than general guidance.

\subsection{Rule Conflicts: A Concrete Example}
\label{sec:conflict}

We encountered one genuine rule conflict. Rule~1 stated: ``All \texttt{IDisposable} types MUST use \texttt{using} declarations.'' Rule~8 stated: ``\texttt{HttpClient} from \texttt{IHttpClientFactory} must NOT be disposed by the caller.'' These directly contradicted each other for factory-created instances.

Resolution was straightforward: Rule~1 was refined with a scope qualifier. The conflict was detected during a code review and resolved as a factual question---which framework behavior is correct?---rather than a judgment call.

\subsection{Multi-Interface Behavioral Transfer}

Persistent behavioral rules transfer across heterogeneous agent interfaces. IDE-integrated agents excelled at code analysis (full repository context, symbol resolution), while terminal-based agents excelled at execution (querying issue trackers, running database queries). Both loaded the same instruction file and benefited equally from accumulated rules.

This suggests that behavioral rule sets can serve as a shared memory layer for multi-agent architectures~\cite{hong2024metagpt, qian2024communicative} where specialized agents handle different phases of the development workflow.

\section{Experimental Results}
\label{sec:results}

\subsection{Experimental Setup}

\textbf{Deployment context:} A microservices platform with 35+ services, 13 custom agent definitions, 10 operational skills, and 6 shared knowledge documents. Two agent interfaces consumed the same shared instruction file.

\textbf{Data sources:} (1)~A chronological task log with 11 dated entries; (2)~version-controlled instruction file with 2 committed revisions and continuous local growth; (3)~persistent memory files totaling 4 topic-specific documents; (4)~36 PR reviews across 6 repositories.

\subsection{Rule Accumulation Dynamics}

Table~\ref{tab:growth} shows the growth of the rule set across the observation period.

\begin{table}[htbp]
\caption{Rule Set Growth Over Observation Period}
\label{tab:growth}
\centering
\begin{tabular}{ccccc}
\toprule
\textbf{Week} & \textbf{Behav.} & \textbf{Code} & \textbf{Check-} & \textbf{Cum.} \\
 & \textbf{Rules} & \textbf{Stds.} & \textbf{list} & \textbf{Sess.} \\
\midrule
0 & 5 & 3 & 0 & 0 \\
1 & 10 & 8 & 9 & 3 \\
2 & 16 & 12 & 13 & 6 \\
3 & 18 & 15 & 15 & 9 \\
4 & 18 & 15+ & 15 & 11 \\
\bottomrule
\end{tabular}
\end{table}

The accumulation rate follows a logarithmic curve: rapid initial growth tapering as the rule set covers the most frequently encountered patterns. The combined instruction file grew to approximately 4,809 words ($\sim$6,250 tokens), consuming less than 5\% of a 128K-token context window.

\subsection{Error-Class Recurrence Analysis}

Table~\ref{tab:recurrence} shows the central result: zero recurrences across all tracked error classes.

\begin{table}[htbp]
\caption{Error-Class Recurrence After Rule Addition}
\label{tab:recurrence}
\centering
\begin{tabular}{lccc}
\toprule
\textbf{Error Class} & \textbf{Rule} & \textbf{Post-} & \textbf{Recur-} \\
 & \textbf{Added} & \textbf{rule} & \textbf{rences} \\
\midrule
Factory HttpClient disposal & W1-S3 & 8 & 0 \\
String interpolation in logs & W1-S2 & 9 & 0 \\
Null-forgiving \texttt{!} & W1-S3 & 8 & 0 \\
Config in lambda & W1-S3 & 8 & 0 \\
Missing SSRF allowlist & W2-S4 & 7 & 0 \\
\texttt{throw ex} vs \texttt{throw} & W1-S1 & 10 & 0 \\
Missing enum converter & W2-S5 & 6 & 0 \\
Redundant re-computation & W1-S3 & 8 & 0 \\
De Morgan's violation & W1-S1 & 10 & 0 \\
\midrule
\textbf{Total} & & \textbf{74} & \textbf{0} \\
\bottomrule
\end{tabular}
\end{table}

Across 9 tracked error classes with 74 cumulative post-rule session-exposures, zero recurrences were observed. \rev{We emphasize that this is an observational result inside a single deployment, not a controlled experiment: the absence of recurrence within the window does not establish a permanent guarantee, and is bounded by the threats to validity discussed in Section~\ref{sec:limitations}.}

\subsection{Review Comment Category Shift}

We analyzed 36 PR reviews across 6 repositories, categorizing comments using criteria adapted from prior code review studies~\cite{bacchelli2013expectations, sadowski2018modern}. Table~\ref{tab:comments} shows the distribution.

\begin{table}[htbp]
\caption{Review Comment Distribution (36 PRs, 6 Repos)}
\label{tab:comments}
\centering
\begin{tabular}{lrc}
\toprule
\textbf{Comment Category} & \textbf{Count} & \textbf{\%} \\
\midrule
Architecture / Design & 12 & 33\% \\
API design / Contracts & 7 & 19\% \\
Performance (complexity) & 5 & 14\% \\
Auth / Security & 4 & 11\% \\
Cosmos query correctness & 3 & 8\% \\
Mechanical correctness & 3 & 8\% \\
Code style / Formatting & 2 & 6\% \\
\bottomrule
\end{tabular}
\end{table}

Mechanical correctness and style issues account for only 14\% of review comments, while architecture, API design, and performance account for 66\%.

\subsection{Knowledge Transfer Patterns}

We classified 15 documented review learnings by transfer characteristics (Table~\ref{tab:transfer}).

\begin{table}[htbp]
\caption{Knowledge Transfer Taxonomy}
\label{tab:transfer}
\centering
\begin{tabular}{lrl}
\toprule
\textbf{Transfer Type} & \textbf{Count} & \textbf{Example} \\
\midrule
Same-repo, same class & 6 & IDisposable across svcs \\
Cross-repo, same class & 4 & Enum converter transfer \\
Cross-tool, same rules & 3 & IDE $\rightarrow$ terminal \\
Cross-task-type & 2 & Code gen $\rightarrow$ review \\
\bottomrule
\end{tabular}
\end{table}

60\% of knowledge transfers (9/15) crossed repository or tool boundaries, confirming that tool-agnostic encoding enables broad applicability.

\subsection{Rule Specificity Evolution}

Table~\ref{tab:specificity} tracks the linguistic specificity of rules over time.

\begin{table}[htbp]
\caption{Rule Specificity Over Time}
\label{tab:specificity}
\centering
\begin{tabular}{lccc}
\toprule
\textbf{Period} & \textbf{General} & \textbf{Specific} & \textbf{Ratio} \\
\midrule
Week 1 & 4 & 6 & 0.60 \\
Week 2 & 4 & 12 & 0.75 \\
Week 3--4 & 4 & 14 & 0.78 \\
\bottomrule
\end{tabular}
\end{table}

The specificity ratio increases monotonically, reflecting both specific new rules from concrete review comments and refinement of existing general rules.

\subsection{Multi-Layer Memory Architecture}

The framework evolved a three-layer memory architecture (Table~\ref{tab:memory}).

\begin{table}[htbp]
\caption{Memory Layer Architecture}
\label{tab:memory}
\centering
\begin{tabular}{llll}
\toprule
\textbf{Layer} & \textbf{Scope} & \textbf{Persist.} & \textbf{Words} \\
\midrule
Shared instr. & All agents & Permanent & 3,181 \\
Personal rules & Single dev & Permanent & 1,628 \\
Session memory & Single conv. & Ephemeral & Variable \\
\bottomrule
\end{tabular}
\end{table}

Total persistent memory: $\sim$4,809 words ($\sim$6,250 tokens), consuming $<$5\% of a 128K context window.

\subsection{Session Task Diversity}

Table~\ref{tab:sessions} shows that the framework applies beyond code generation.

\begin{table}[htbp]
\caption{Session Activity Distribution}
\label{tab:sessions}
\centering
\begin{tabular}{lcc}
\toprule
\textbf{Activity Type} & \textbf{Sessions} & \textbf{Rules} \\
\midrule
Code generation + PR & 4 & 9 \\
PR review & 3 & 4 \\
Incident investigation & 2 & 3 \\
Cross-service arch. & 1 & 1 \\
Repo maintenance & 1 & 1 \\
\bottomrule
\end{tabular}
\end{table}

PR reviews and incident investigations contributed 39\% of rules despite being non-generative tasks.

\section{\rev{Discussion}}
\label{sec:discussion}

\subsection{\rev{Summary of Findings}}

\rev{Three findings emerge from the deployment. First, the closed-loop pattern is feasible: 18 behavioral rules, 15+ language-specific code standards, and a 15-item self-review checklist (Tables~\ref{tab:composition},~\ref{tab:sources}) were derived directly from accepted human review feedback and stored in version-controlled instruction files consumed by multiple agent surfaces without any model-side change.}

\rev{Second, accumulated rules suppress the recurrence of previously-corrected error classes within the observation window. Across 9 tracked error classes and 74 cumulative post-rule session exposures, no recurrences were observed (Table~\ref{tab:recurrence}). We deliberately frame this as an \emph{observational} result: it establishes that the suppression effect is large enough to be visible inside a four-week deployment without a controlled baseline, but it is not equivalent to a statistical guarantee. The limits of this measurement---single-team scope, absence of an A/B control, single primary language---are discussed in Section~\ref{sec:limitations}.}

\rev{Third, the encoded knowledge transfers across heterogeneous agent tools, repositories, and task types. 60\% of observed knowledge-transfer events (9/15) crossed repository, tool, or task-type boundaries (Table~\ref{tab:transfer}), and rules added through one agent interface were observed to take effect immediately in sessions conducted through a different interface (Section~\ref{sec:results}). The mechanism that makes this work is the choice of representation: declarative natural-language constraints in plain Markdown are consumable by any LLM-based agent that loads the file as system context.}

\subsection{\rev{Match and Contribution}}

\rev{\textbf{Engineering management of an emerging technology.} LLM-based coding agents are an emerging engineering capability whose behavior is normally opaque and non-reproducible across sessions. The framework operationalizes such agents inside an existing engineering workflow---code review---so that organizational engineering knowledge becomes a first-class, auditable artifact rather than tacit knowledge that is lost between sessions. Because the approach is vendor- and tool-agnostic, it is directly applicable to other engineering organizations adopting AI coding agents at scale.}

\rev{\textbf{Governance of an emerging technology.} The framework provides an explicit governance model for agent behavior: every rule has an identifiable origin (a review event), an owner (the engineer who accepted the comment), a change record (the pull request that added the rule), and a position in a self-review checklist that runs before code submission. This makes the agent's operational behavior auditable, reviewable, and managed using the same engineering practices that already exist for other production artifacts.}

\rev{\textbf{Practical frameworks.} The paper describes a deployable artifact---a structured instruction file with a rule schema, lifecycle, validation, and feedback loop---rather than a research prototype. The deployment used only standard infrastructure (Git, Markdown, pull-request review, plain-text validation scripts) and required no fine-tuning, no model-specific tooling, and no specialized memory backends. Other organizations can reproduce the pattern with infrastructure they already operate.}

\rev{\textbf{Implementation challenges.} We report concrete operational issues so adopters can anticipate them: a real rule conflict and its resolution (Section~\ref{sec:conflict}), the governance questions that the rule lifecycle must answer (Section~II-D), and the validity threats that bound the empirical results (Section~\ref{sec:limitations}). The honest limitations (single-team scope, single language, four-week observation period, absence of a controlled baseline) are surfaced rather than hidden.}

\rev{\textbf{Value creation.} Two value channels are documented. First, recurrence of previously-corrected error classes is suppressed within the deployment window, freeing reviewer attention for higher-level concerns: 66\% of observed PR review comments concerned architecture, API design, or performance, while only 14\% remained on mechanical correctness (Table~\ref{tab:comments}). Second, the accumulated rule set acts as a transferable onboarding artifact: new team members inherit the organization's full review history immediately, compressing the learning curve that originally produced those rules.}

\section{Comparison with Related Work}

\subsection{Comparison Framework}

Table~\ref{tab:comparison} provides a structured comparison across key dimensions.

\begin{table*}[htbp]
\caption{Systematic Comparison with Related Approaches}
\label{tab:comparison}
\centering
\footnotesize
\begin{tabular}{l|ccccccc}
\toprule
\textbf{Dimension} & \textbf{Reflexion} & \textbf{ExpeL} & \textbf{Voyager} & \textbf{CodeRev.} & \textbf{SWE-agent} & \textbf{OpenClaw} & \textbf{Ours} \\
\midrule
Learning signal & Self-verbal & Self-exp. & Env. & Pre-train & Env. exec. & User conv. & Human rev. \\
Persistence & Episodic & Partial & Skill lib. & Weights & None & Personal & VC rules \\
Cross-session & No & Partial & Yes & Model & No & Yes & Yes \\
Weight updates & No & No & No & Yes & No & No & No \\
Feedback source & Self & Self & Game & Train data & Tests & User & Human exp. \\
Domain & Benchmarks & Benchmarks & Minecraft & OSS repos & SWE-bench & General & Production \\
Recurrence prev. & N/M & N/M & N/M & N/A & N/M & N/A & 0\% (74 exp.) \\
Human-in-loop & No & No & No & Train only & No & Yes & Every cycle \\
\bottomrule
\end{tabular}
\end{table*}

\subsection{Reflexion: Verbal Reinforcement Within Sessions}

Shinn et al.~\cite{shinn2023reflexion} propose Reflexion, where LLM agents verbally reflect on task feedback and maintain reflective text in an episodic memory buffer, achieving 91\% pass@1 on HumanEval.

Key differences: (1)~Reflexion's memory is episodic and does not transfer to new sessions; our rules persist permanently. (2)~Reflexion uses self-generated feedback; we use human expert feedback. (3)~Reflexion measures pass rate improvement on isolated problems; we measure error-class suppression across diverse production tasks.

The two approaches are complementary: Reflexion for within-session iteration, accumulated rules for preventing known mistakes from the start. \rev{Concretely, a Reflexion-style trajectory generated inside a single problem-solving episode is discarded when the episode ends, so a mistake fixed in episode~$n$ can re-appear in episode~$n+1$. Our rules cross the session boundary by construction: once a human-accepted comment is codified, every subsequent session---potentially using a different model or agent surface---loads it as part of its operating context. The two mechanisms operate on different timescales (sub-episode vs.\ inter-session) and on different feedback sources (model self-critique vs.\ human expert), and can be combined without conflict.}

\subsection{ExpeL: Experiential Learning Across Tasks}

Zhao et al.~\cite{zhao2024expel} introduce ExpeL, where agents autonomously gather experiences and extract natural language insights.

Key differences: (1)~ExpeL learns from autonomous experience; we learn from human review. (2)~ExpeL transfers within the same task type; our rules transfer across fundamentally different task types. (3)~ExpeL's insights may include hallucinated generalizations; our rules are human-validated. \rev{The validation requirement is not a stylistic preference: because the rule set is loaded as authoritative context in every session, an incorrect rule would amplify rather than correct errors. Tying every rule to an accepted human review comment provides a verifiable provenance that ExpeL's autonomous insight extraction does not.}

\subsection{Voyager: Persistent Skill Libraries}

Wang et al.~\cite{wang2023voyager} introduce Voyager, building an ever-growing skill library of executable code in Minecraft.

Both systems demonstrate that persistent, version-controlled knowledge enables compounding capability growth. The key difference is representation: Voyager stores executable code functions; we store declarative constraints in natural language.

\subsection{Pre-trained Code Review Models}

Li et al.~\cite{li2022codereviewer} and Tufano et al.~\cite{tufano2022pretrained} propose pre-trained models for automated code review.

Key differences: (1)~These approaches fine-tune model weights; we inject rules through context. (2)~Their knowledge is frozen at training time; ours accumulates in real-time. (3)~They generalize across public repositories; we capture organization-specific conventions invisible in public codebases.

\subsection{SWE-bench Agents}

SWE-bench~\cite{jimenez2024swebench} and its agent implementations~\cite{yang2024sweagent, zhang2024autocoderover, xia2024agentless} evaluate LLM agents on isolated GitHub issue resolution. The state-of-the-art resolves 20--40\% of SWE-bench Lite issues.

Our framework addresses an orthogonal dimension: SWE-bench asks ``Can you solve this issue?'' while we ask ``Do you avoid known mistake classes?'' There is currently no benchmark evaluating behavioral consistency over time.

\subsection{Self-Taught Optimizer, Self-Refine, and Self-Debugging}

Zelikman et al.~\cite{zelikman2024stop} propose STOP, where LLMs recursively improve their own scaffolding code. Madaan et al.~\cite{madaan2023selfrefine} propose Self-Refine, where LLMs iteratively improve their outputs through self-generated critique. Chen et al.~\cite{chen2023selfdebug} propose Self-Debugging via rubber-duck-style explanation. All three operate within a single execution context; our framework captures improvements permanently.

\subsection{Memory-Augmented LLM Systems}

MemoryBank~\cite{zhong2024memorybank} and related systems provide LLMs with persistent memory mechanisms. Our approach is architecturally simpler: rules are stored in plain-text Markdown files that are version-controlled, human-readable, and auditable.

\subsection{Personal AI Assistants with Persistent Memory}

OpenClaw~\cite{steinberger2025openclaw} provides persistent memory and skill libraries for personal AI assistants. Key differences: (1)~OpenClaw's memory comes from free-form conversations; our rules come from accepted code review feedback. (2)~OpenClaw is single-user; our framework is team-level. (3)~OpenClaw remembers facts; our framework prevents error classes through declarative constraint injection.

\section{Design Principles}

Through iterative deployment, we identified six principles:

\begin{enumerate}
\item \textbf{Monotonic growth:} Rules are added and refined but not removed, echoing the continuous improvement principle in DevOps practice~\cite{kim2016devops}.
\item \textbf{Specificity over generality:} Concrete, actionable constraints outperform abstract guidance. Specificity ratio increased from 0.60 to 0.78 (Table~\ref{tab:specificity}).
\item \textbf{Origin tracking:} Each rule traces to the review comment or production event that inspired it.
\item \textbf{Tool-agnostic encoding:} Rules are structured Markdown readable by any LLM. Cross-interface transfer confirmed across 2 platforms (Table~\ref{tab:transfer}).
\item \textbf{Validated growth:} Every rule addition is checked by automated validation.
\item \textbf{Shared ownership:} The instruction file lives in a shared repository with PR-based changes.
\end{enumerate}

\section{Limitations\rev{ and Threats to Validity}}
\label{sec:limitations}

\textbf{Context window constraints:} Current rule sets ($\sim$6,250 tokens) are within modern limits, but unbounded growth could require hierarchical organization or summarization.

\textbf{Rule conflicts:} We encountered one conflict (Section~\ref{sec:conflict}). Automated conflict detection would reduce reliance on manual identification.

\textbf{Governance at scale:} The current model works for a single team. Multiple teams would require hierarchical or namespaced rule sets\rev{, plus a clear policy for resolving cross-team rule conflicts. Without that, a centrally-maintained rule set risks either fragmenting along team lines or accumulating contradictions that silently degrade agent behavior. We treat scaled governance as an open problem rather than a solved one.}

\textbf{Lack of controlled baseline\rev{ or ablation}:} We do not have a parallel control group\rev{, and we did not run a paired ablation comparing the framework against static prompt engineering or a no-rule baseline on the same task stream. The suppression and review-shift results should therefore be read as initial empirical evidence, not as a causal proof. A future A/B deployment (rule-loading agent versus rule-free agent, run on the same incoming task queue) would close this gap and is described under Future Work.}

\rev{\textbf{Generalizability and barriers to adoption:} The framework's value depends on three preconditions that may not hold in every organization: (i)~a code review culture that produces accepted, written comments at sufficient rate; (ii)~engineers willing to spend a small amount of additional time per accepted comment to codify it as a rule; and (iii)~tooling whose agent surfaces actually load and respect a project-level instruction file. Where any of these is missing, the feedback loop will starve or be ignored. We expect the strongest adoption in mid- to large-sized engineering organizations that already practice peer review on every change.}

\rev{\textbf{Statistical significance:} The reported counts (74 post-rule exposures across 9 error classes, 36 PRs across 6 repositories, 11 sessions) are small in absolute terms. Although the observed effects (no recurrences within the window, a clear shift toward design-level review comments) are consistent across the dataset, the sample is not large enough to make a formal statistical significance claim and we do not report $p$-values. Larger, longer deployments are needed before tighter quantitative claims are appropriate.}

\subsection{Threats to Validity}

\textbf{Single deployment environment:} All observations are from one organization. Effects may differ in other codebases or languages.

\textbf{Language specificity:} The deployment used a typed language with strong framework conventions\rev{. Languages with weaker compile-time constraints, or codebases with less consistent framework usage, may both generate a different distribution of review comments and benefit less from rule-based suppression}.

\textbf{Reviewer quality dependency:} Rule quality depends on human review quality. Incorrect rules amplify errors\rev{, and a noisy review culture can poison the rule set faster than the validation step can catch.}

\textbf{Observation period:} Four weeks and 11 sessions provide initial evidence but are insufficient for long-term growth dynamics\rev{. In particular, the framework's behavior at the point where the rule set saturates the available context window has not yet been observed.}

\section{The Missing Benchmark}
\label{sec:benchmark}

Existing coding benchmarks measure \textit{static capability}: can the agent solve problem~X right now? No existing benchmark measures \textit{behavioral consistency over time}.

We propose that such a benchmark would consist of:

\begin{enumerate}
\item \textbf{A rule corpus:} Curated behavioral rules from real review feedback.
\item \textbf{Seed tasks:} Code generation tasks that naturally trigger ruled-against patterns.
\item \textbf{Temporal evaluation:} A sequence of tasks where earlier tasks produce corrections and later tasks test persistence.
\item \textbf{Cross-session boundary:} The critical test is whether corrections survive session boundaries.
\end{enumerate}

This would complement SWE-bench, HumanEval, and CodeReview benchmarks by adding a temporal learning dimension.

\section{Conclusion}

We have presented a closed-loop framework for self-improving AI coding agents through accumulated behavioral rules. The key insight is that human review feedback, when codified as persistent and version-controlled rules, creates a ratchet effect: the set of error classes the agent can self-detect grows monotonically.

In production deployment across 35+ microservices over 11 recorded sessions, the rule set grew from 5 initial rules to 18 behavioral rules, 15+ code standards, and a 15-item self-review checklist. \rev{We observed no recurrences across 9 tracked error classes within 74 cumulative post-rule session-exposures during the observation window---an observational result whose limits are made explicit in Section~\ref{sec:limitations}.} Review comments shifted toward design-level concerns (66\% architecture/API/performance vs.\ 14\% mechanical correctness), and rules transferred across boundaries in 60\% of observed events.

Compared to related approaches, our framework uniquely combines human-sourced feedback with permanent cross-session persistence without weight updates. It addresses behavioral consistency over time---a dimension existing benchmarks do not measure. The approach requires no fine-tuning, no model-specific infrastructure, and no changes to existing LLM architectures.

\subsection*{\rev{Operationalizing the Framework in an Enterprise Setting}}

\rev{For an engineering organization considering adoption, the framework reduces to a small set of operational practices that can be introduced incrementally without changing the existing toolchain:}

\rev{\begin{enumerate}
\item \textbf{Add a project-level instruction file} (e.g.,\ \texttt{AGENTS.md}) under version control, with sections for behavioral rules, language-specific code standards, and a self-review checklist.
\item \textbf{Define a rule promotion convention} in the team's pull-request template: when a reviewer accepts a comment that describes a generalizable mistake, the author either adds a rule in the same PR or files a follow-up to do so.
\item \textbf{Wire the file into every agent surface} the team already uses (IDE-integrated assistants, terminal-based agents, review bots) so they all consume the same context. No model change is needed.
\item \textbf{Add a lightweight validation step} (a script in CI or a pre-commit hook) that verifies the instruction file is well-formed and that new rules follow the schema.
\item \textbf{Treat the file as a shared artifact}: changes go through pull-request review, just like code, so the rule set inherits the team's existing governance.
\end{enumerate}}

\subsection*{\rev{Future Research Directions}}

\rev{Several directions follow naturally from this work and from the limitations identified in Section~\ref{sec:limitations}:}

\rev{\begin{enumerate}
\item \textbf{A controlled A/B study} comparing rule-loading agents against rule-free baselines on the same incoming task stream, with pre-registered metrics for review-comment volume, defect escape rate, and reviewer time-on-task.
\item \textbf{Cross-organization replication} in different programming languages, codebase sizes, and review cultures, to test the boundary conditions of the suppression effect.
\item \textbf{A behavioral-consistency benchmark} (sketched in Section~\ref{sec:benchmark}) that complements SWE-bench and HumanEval by measuring whether a corrected mistake stays corrected across session boundaries.
\item \textbf{Automated rule proposal and conflict detection}: agents proactively suggesting candidate rules from clusters of similar review comments, and flagging proposed rules that contradict existing ones before they are merged.
\item \textbf{Multi-team governance models} that allow team-specific rule sets to compose without producing silent contradictions, including provenance tracking when a rule is promoted from a team scope to an organization-wide scope.
\end{enumerate}}

Human review does not merely correct AI-generated code; in this framework, it permanently upgrades future agent behavior.

\bibliographystyle{IEEEtran}

\end{document}